\documentclass[conference]{IEEEtran}
\IEEEoverridecommandlockouts

\usepackage[T1]{fontenc}
\usepackage{cite}
\usepackage{amsmath,amssymb,amsfonts}
\usepackage{algorithmic}
\usepackage[dvipsnames]{xcolor}
\usepackage{graphicx}
\usepackage{textcomp}
\usepackage{booktabs}
\usepackage{multirow}
\usepackage{hyperref}
\usepackage{pifont}
\usepackage{comment}

\begin{document}

\title{Guardians of DNS Integrity: A Remote Method for Identifying DNSSEC Validators Across the Internet}

\author{\IEEEauthorblockN{Yevheniya Nosyk, Maciej Korczy\'nski, Andrzej Duda}

\IEEEauthorblockA{Univ.~Grenoble Alpes, CNRS, Grenoble
INP, LIG,
38000 Grenoble, France   %
\\ Email: firstname.lastname@univ-grenoble-alpes.fr
}
}

\maketitle

\begin{abstract}
DNS Security Extensions (DNSSEC) provide the most effective way to fight DNS cache poisoning attacks. Yet, very few DNS resolvers perform DNSSEC validation. Identifying such systems is non-trivial and the existing methods are not suitable for Internet-scale measurements. In this paper, we propose a novel remote technique for identifying DNSSEC-validating resolvers. The proposed method consists of two steps. In the first step, we identify open resolvers by scanning 3.1 billion end hosts and request every non-forwarder to resolve one correct and seven deliberately misconfigured domains. We then build a classifier that discriminates validators from non-validators based on query patterns and DNS response codes. We find that while most open resolvers are DNSSEC-enabled, less than 18\% in IPv4 (38\% in IPv6) validate received responses. In the second step, we remotely identify closed non-forwarders in networks that do not have inbound Source Address Validation (SAV) in place. Using the classifier built in step one, we identify 37.4\% IPv4 (42.9\% IPv6) closed DNSSEC validators and cross-validate the results using RIPE Atlas probes. Finally, we show that the discovered (non)-validators actively send requests to DNS root servers, suggesting that we deal with operational recursive resolvers rather than misconfigured machines.
\end{abstract}


\section{Introduction}

All communications on the Internet rely on the Domain Name System (DNS) that maps IP addresses to domain names, which are further used to access various resources on the Internet. The original DNS implementation~\cite{rfc1035} was not designed with security in mind and did not provide any means for data authentication. So, DNS resolvers cannot guarantee that the received responses are genuine. 

The DNS \emph{cache poisoning} attack~\cite{kaminsky, cache_injections, patched_dns,fragmentation_leaking,vulnerable_delegation,socket_overloading,fragmentation_poisonous,collaborative,wrinkle,corrupted} leverages this vulnerability. An attacker injects a bogus DNS entry into the recursive resolver's cache before the genuine reply arrives from the authoritative nameserver. Once cached, it will be returned in response to future client requests directly. The reasons for an attacker to poison the cache may include modifying the glue records associated with a domain name or redirecting traffic from a legitimate domain by inserting a \texttt{\small CNAME} record~\cite{cache_injections}. When launched remotely, cache poisoning can only target \emph{open resolvers}. However, recent studies showed that \emph{closed resolvers} can also be reached from outside the trusted range of IP addresses if the source IP of the attacker packet is forged~\cite{korczyski2020dont, korczyski2020closed, anrw}. While cache poisoning may be the final goal of an attacker, it may also be an intermediate step towards other attacks or various forms of DNS manipulations~\cite{manipulation}. 

One way to address the problem of cache poisoning is to perform source port and query ID randomization~\cite{rfc5452}. However, a more comprehensive approach is to deploy DNS Security Extensions (DNSSEC)~\cite{rfc4033} that support data authentication and integrity using public-key cryptography. The DNS zone administrator generates public/private key pairs to digitally sign the resource records and adds the signatures, along with the public keys and their fingerprints, to the zone files using new resource records: \texttt{\small RRSIG}, \texttt{\small DNSKEY}, \texttt{\small NSEC(3)} and \texttt{\small DS}~\cite{rfc4034}. The \emph{validating recursive resolver} (or just \emph{validator}) retrieves and analyzes the DNSSEC-related resource records to verify the integrity of the received query response~\cite{rfc4035}.

For DNSSEC to be effective, it needs to be widely deployed: DNS zones have to be correctly signed and recursive resolvers have to validate all the query responses they receive. Previous work has shown that the authoritative nameserver side of DNSSEC is highly mismanaged and the population of signed domains is relatively low~\cite{longitudinal}. Only a few registrars attempted to facilitate DNSSEC deployment for their customers~\cite{registrars}.

While measuring the extent of DNSSEC deployment is possible through inspection of relevant records, enumerating validating DNS resolvers remains a huge challenge. Several researchers proposed a few methods such as sending requests to controlled domains from geographically distributed vantage points~\cite{longitudinal}, triggering clients to resolve certain domains~\cite{practical_impact,measuring_occurence,check_repeat}, and passively analyzing logs on authoritative nameservers of top-level domains (\texttt{\small .jp}, \texttt{\small .org})~\cite{wild,counting}. However, the proposed approaches do not scale for the measurements of the global Internet, require paid services, or access to privileged data.

In this paper, we propose a novel active method for identifying DNS resolvers that perform DNSSEC validation in the IPv4 and IPv6 address spaces at the scale of the Internet. It overcomes the two greatest challenges involved in trying to discover validating DNS resolvers out there: it does not require access to any remote vantage points nor to restricted data. The proposed method consists of two steps. We first actively enumerate \emph{open resolvers} by scanning the whole routable IPv4 space~\cite{routeviews} and a targeted list of IPv6 addresses~\cite{hitlist}. We send requests to 3.1B target IP addresses to resolve unique domain names under our control. Then, we probe every non-forwarding open resolver with a set of eight domains: one with correct DNSSEC configuration and seven others deliberately misconfigured. We analyze query patterns of open resolvers on our two-level signed authoritative DNS zone infrastructure and the responses received on our scanner to build a classifier that effectively distinguishes validating from not validating resolvers. The second step extends the technique proposed by Korczy\'nski et al.~\cite{korczyski2020dont} to identify \emph{closed resolvers}. More specifically, we again probe 3.1B hosts with unique domain names under our control and observe the requests on our authoritative nameservers (as we spoof source addresses, we cannot receive the responses returned by the closed resolvers). By classifying the query patterns of closed resolvers using the models build in the first step, we identify the closed resolvers as DNSSEC validators only based on the request pattern on our authoritative nameservers.

\begin{figure}[t]
    \centering
    \includegraphics[width=0.475\textwidth]{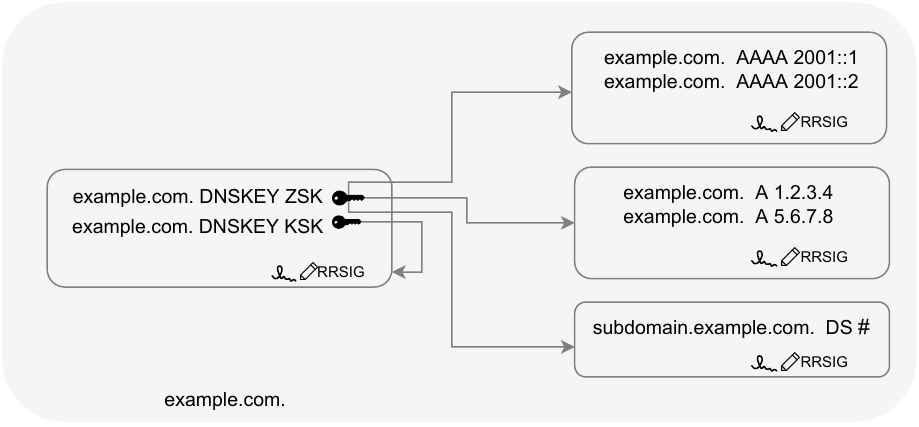}
    \caption{Zone structure of a signed domain \texttt{example.com.} Each rectangle represents a resource record set (RRset). Three of them (\texttt{A},\texttt{AAAA}, and \texttt{DS}) are signed with a Zone Signing Key (ZSK), while \texttt{DNSKEY} RRset is signed with a Key Signing Key (KSK). Note that the \texttt{DS} record was generated by the subdomain of \texttt{example.com}.}
    \label{signing}
\end{figure}

Our Internet-wide scans reveal 6.9M open and closed IPv4/IPv6 DNS resolvers - several times more than the numbers obtained by previous work. As we narrowed down our analysis to DNSSEC-enabled non-forwarders only,  we evaluated DNSSEC validation policies of 32.2K open and 400K closed resolvers, all originating from 224 countries and 21.9K organizations (31.2\% of all the routable autonomous systems). In particular, we found that while most of the tested non-forwarding open resolvers (78.8\% IPv4 and 92.6\% IPv6) claim to be DNSSEC-enabled (by setting the \texttt{\small DO} bit), less than 18\% in IPv4 and 38\% in IPv6 of them actually validate received responses. Interestingly, significantly more closed resolvers are revealed to be validators - 37.4\% IPv4 and 42.9\% IPv6. We obtained the ground truth data for 123 of those resolvers using RIPE Atlas measurement network~\cite{ripe_atlas} and confirmed that our passive technique correctly classified 91\% of them. Finally, we inspected around 301B queries on the DNS root servers and found that 7\% of them (more than 21B) were initiated by resolvers that we identified. It suggests that those are operational recursive resolvers with end clients/systems behind them. The presented classification model was created as a result of active measurements but it can also be used ``in the wild'' on passive traces from authoritative nameservers.

The rest of the paper is organized as follows. Section~\ref{background_related} provides the background on DNSSEC and analyzes related work on evaluating DNSSEC validation. In Section~\ref{compliance}, we test several DNS server implementations and show how they handle DNSSEC-related queries. Section~\ref{active} describes the first step of the method and presents the results of identifying open validating resolvers. Section~\ref{passive} introduces the second step of the method to find validating closed resolvers. We analyze the ethical impact of our study in Section~\ref{ethics}. Section~\ref{conclusions} concludes the paper and gives some thoughts for future work. 

\section{Background on DNSSEC and Related Work\label{background_related}}

This section provides the background on DNSSEC and describes related work on the state of DNSSEC validation.

\begin{table*}[t]
    \caption{Existing methods on identifying DNSSEC validators} 
    \label{related_work} 
    \scriptsize
    \centering 
    \begin{tabular}{lcccccc}
        \toprule
            \textbf{Method} & \textbf{Remote} & \textbf{Paid} & \textbf{Uses privileged data} & \textbf{Duration} &  \textbf{Coverage} \\
        \midrule
            Chung et al.~\cite{longitudinal} &  No & Yes & No & 13 days & 59,513 resolvers / 403,355 end clients\\
            Lian et al.~\cite{practical_impact} & Yes & Yes & No & 7 days & 35,010 resolvers / 529,294 end clients \\
            Wander and Weis~\cite{measuring_occurence} & Yes & No & No & 7 months  & 98,179 end clients\\
            Yu et al.~\cite{check_repeat} & Yes & No & No & 36 days  & 49,488 resolvers\\
            Gu{\dh}mundsson and Crocker~\cite{wild} & Yes & No & Yes &  50 / 50 minutes  & 676,599 / 573,773 resolvers\\
            Fukuda et al.~\cite{counting} & Yes & No & Yes & 2 / 2 / 2 days & 2,150,958 / 2,081,826 / 1,904,610 resolvers\\
            Our method & Yes & No & No & 15 days & 6,880,023 resolvers\\
        \bottomrule
    \end{tabular}
\end{table*}

\begin{figure}[t]
  \centering
   \includegraphics[width=0.5\textwidth]{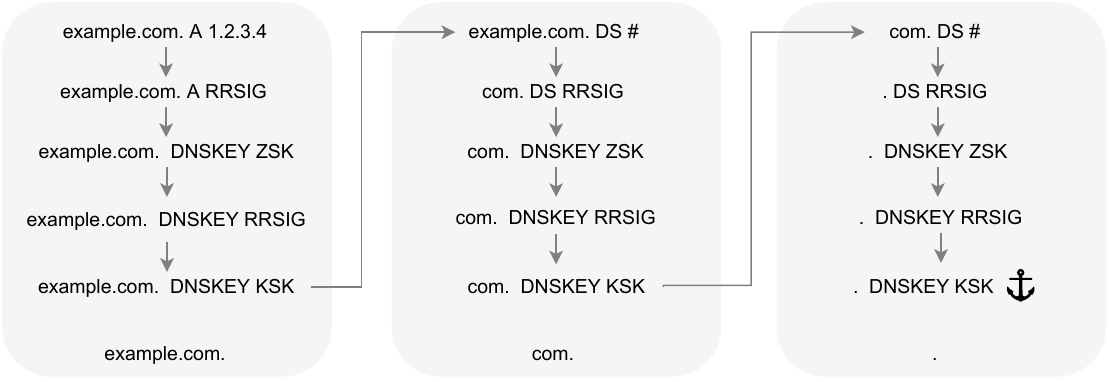}
  \caption{A sequence of queries performed by a validating recursive resolver. It forms a chain of trust from the initially queried \texttt{A} record of \texttt{example.com} up until the trust anchor (KSK of the root).}
  \label{validation}
\end{figure}

\subsection{Background on DNSSEC\label{background}}

RFC 4033, 4034, and 4035~\cite{rfc4033,rfc4034,rfc4035} define the DNSSEC operation. Two key concepts are \emph{zone signing} and \emph{response validation}.

A DNS zone is cryptographically signed using a private~/ public key pair. Figure~\ref{signing} presents an example of a signed zone (\texttt{\small example.com}). Although not strictly required by the standard, each zone usually has two key pairs: the Key Signing Key (KSK) and the Zone Signing Key (ZSK). The corresponding public keys are added to the zone in the form of the \texttt{\small DNSKEY} resource records. The private KSK signs the \texttt{\small DNSKEY} resource record set (RRset) and generates the \texttt{\small RRSIG} signature. The RRsets containing other names in the zone are similarly signed with the private ZSK key. The \texttt{\small DS} record is published to the parent zone of the signed domain and contains the digest of the KSK public key. Zone signing and publishing all necessary records enable integrity validation but it is up to recursive resolvers to check the signatures and validate all the received responses.

Authoritative nameservers serving signed zones should include signatures and keys in query responses if they fit the packet size. However, they significantly increase the response size and total resolution overhead. Recursive resolvers explicitly declare their support for DNSSEC by setting the ``DNSSEC OK'' (\texttt{\small DO}) bit. Authoritative nameservers only include DNSSEC-related records in response to the queries with the \texttt{\small DO} bit set.

DNSSEC-validating recursive resolvers should computationally prove that the received signatures are valid using public keys and hashes published in zones. They also verify the keys by establishing the chain of trust from the trust anchor to a given zone. Figure~\ref{validation} shows the sequence of DNS requests generated by a validating recursive resolver. Once the resolver receives the response to its initial query along with the signature, it requests the public ZSK key of \texttt{\small example.com} to verify the signature validity. It then contacts the parent zone (\texttt{\small .com}) and retrieves the \texttt{\small DS} record of \texttt{\small example.com}, a hash of the \texttt{\small example.com} public KSK key. This process continues up the domain name tree until the recursive resolver reaches the root zone. Every validating resolver is configured with one or more trust anchors. By default, it refers to the root zone's KSK or its hash. The validator compares its configured trust anchors with those returned by the root. If they match, validation ends successfully and the client receives the query response with the Authenticated Data (\texttt{\small AD}) bit set. Should the validation fail for some reason, the resolver returns a generic \texttt{\small SERVFAIL} error. Querying a resolver with a set of correctly signed and misconfigured domains, combined with examining the response codes, gives an insight into whether the tested resolver is DNSSEC-validating.

\subsection{DNSSEC Validation}

Researchers elaborated various methods to test whether recursive resolvers validate the received responses. Table~\ref{related_work} summarizes related work in the field. Each method can be characterized in different ways: remote techniques do not need vantage points to reach resolvers, paid methods require, for example, purchasing an ad space, privileged data may refer to query traces from TLD nameservers, not available to the general public. Finally, we take into account the duration of the experiment and the coverage in terms of recursive resolvers or end clients.

Chung et al.~\cite{longitudinal} deployed one correctly signed and several misconfigured domains and leveraged the paid Luminati HTTP/S proxy service to study the client-side DNSSEC behavior. Out of 59.5K tested resolvers with 403K clients behind, only 543 resolvers detected all the misconfigurations. The method uses a paid service and requires vantage points limited to the geographical locations of Luminati servers.

Lian et al.~\cite{practical_impact} used an online advertisement network with a specifically crafted advertisement that retrieves images hosted on controlled domains. They found that only less than 3\% of clients were protected from bogus DNS zones. Wander and Weis~\cite{measuring_occurence} deployed their client-side tests in several places, including autosurf websites. They found a slightly higher ratio of validating clients (around 4.8\%). These two methods, once again, rely on paid services to analyze resolvers used by end clients.

Yu et al.~\cite{check_repeat} leveraged the Web Proxy Auto-Discovery (WPAD) protocol to collect browser-initiated DNS queries. They relied on the observation that whenever one of the authoritative nameservers returns a bogus reply, a validating client attempts to contact another nameserver. Such repeated queries indicate that the contacting resolver is likely to be validating. They analyzed 49.5K unique resolvers of which 2.4K were validators. This method does not actively probe DNS resolvers and is thus, limited in coverage.

Gu{\dh}mundsson and Crocker~\cite{wild} analyzed queries traced at the \texttt{\small .org} zone. They considered passively observed DNSSEC-related queries as evidence that the contacting resolver is a validator but have not systematically analyzed and proven this hypothesis. The analysis of two groups of query logs detected 4.7K and 5.6K validators. Fukuda et al.~\cite{counting} computed the ratio of the passively observed \texttt{\small DS} queries to all the queries arriving from a given IP address to the \texttt{\small .jp} zone. They acquired the ground truth by actively querying open resolvers and observing the query patterns. The number of potential validators varied significantly in each of the three datasets. Such passive methods only have a limited view of the DNS resolver landscape, as the analysis is restricted to several nameserver traces.

In our experiments, we have analyzed 6.9M recursive resolvers - several times more than in previous methods. Our approach is entirely remote, it does not require access to privileged data (such as query traces from TLD nameservers) or any paid service. We have actively queried 3.1B IPv4 and IPv6 addresses. Therefore, the coverage is much more extended compared, for example, to examining query traces limited in time.

\begin{table*}[t]
    \caption{The implementation of DNSSEC validation by DNS resolver vendors} 
    \label{validator_properties} 
    \scriptsize
    \centering 
    \begin{tabular}{llccccc}
        \toprule
            \multirow{2}{*}{\textbf{Vendor}} & \multirow{2}{*}{\textbf{Version}} & \textbf{DNSSEC validator} & \textbf{Recognizes} & \textbf{Retrieves} & \textbf{Retrieves} & \textbf{Retrieves}\\
            & & \textbf{out-of-the-box} & \textbf{misconfigurations} & \textbf{child \texttt{DNSKEY}} & \textbf{parent \texttt{DNSKEY}} & \textbf{child \texttt{DS}}\\
        \midrule
            BIND9 & 9.16.1-Ubuntu & {\color{LimeGreen}\ding{52}} & {\color{LimeGreen}\ding{52}} &{\color{LimeGreen}\ding{52}} & {\color{LimeGreen}\ding{52}} & {\color{Bittersweet}\ding{56}}\\
            Unbound & 1.9.4 & {\color{LimeGreen}\ding{52}} & {\color{LimeGreen}\ding{52}} & {\color{LimeGreen}\ding{52}} &  {\color{LimeGreen}\ding{52}} & {\color{Bittersweet}\ding{56}} \\
            Knot Resolver & 5.4.4 & {\color{LimeGreen}\ding{52}} & {\color{LimeGreen}\ding{52}} &{\color{LimeGreen}\ding{52}} & {\color{LimeGreen}\ding{52}} & {\color{Bittersweet}\ding{56}} \\
            PDNS-Recursor & 4.2.1-1build2 &  {\color{Bittersweet}\ding{56}} (supported since v4.5.0) & {\color{LimeGreen}\ding{52}} & {\color{LimeGreen}\ding{52}} & {\color{LimeGreen}\ding{52}} & {\color{LimeGreen}\ding{52}} \\
            Microsoft DNS & 10.0.20348 & {\color{Bittersweet}\ding{56}} (need to activate trust anchors) & {\color{LimeGreen}\ding{52}} &  {\color{LimeGreen}\ding{52}} & {\color{LimeGreen}\ding{52}} & {\color{Bittersweet}\ding{56}} \\
        \bottomrule
    \end{tabular}
\end{table*}

\section{Evaluating Compliance with the DNSSEC RFCs\label{compliance}}

We start with preliminary experiments to choose the right DNS software supporting DNSSEC in our testing environment. We report on the tests of five authoritative nameserver and 5 recursive resolver implementations to see how they implement DNSSEC standard. We have set up our experiments on Ubuntu 18.04 LTS and Windows Server 2019 Base virtual machines. We have installed the latest packaged versions of the following DNS server software: BIND9~\cite{bind}, Microsoft DNS~\cite{microsoft}, NSD~\cite{nsd} (authoritative-only), PowerDNS~\cite{pdns}, Knot DNS~\cite{knot-dns} (authoritative-only), Knot Resolver~\cite{knot-resolver} (recursive-only), and Unbound~\cite{unbound} (recursive-only).

\subsection{Authoritative Nameservers\label{auth_nameserver}}

The DNSSEC standard~\cite{rfc4035} states that all the query responses sent out by an authoritative nameserver to a DNSSEC-enabled resolver should be accompanied by corresponding \texttt{\small RRSIG} signatures. The signature must verify the requested resource record, provided the recursive resolver is willing to do the necessary checks. In case the authoritative nameserver cannot provide the response (e.g., the requested domain name or resource record do not exist), an \texttt{\small NSEC}/\texttt{\small NSEC3} resource record is returned as a proof of  non-existence. If the returned response does not directly answer the query but rather refers to the child zone, the KSK hash of the child zone (\texttt{\small DS}) is also to be included. The \texttt{\small DNSKEY} key however, may only be included in response to the \texttt{\small SOA} or \texttt{\small NS} queries arriving at the root of the zone (zone apex).

For each tested nameserver implementation, we set up a two-level DNS zone infrastructure (\texttt{\small subdomain.test.com} and \texttt{\small test.com}) on two distinct machines. Both zones are correctly signed. We then deploy a validating recursive resolver running BIND9 that we use to issue \texttt{\small A} requests for \texttt{\small subdomain.test.com}. Our experiment confirms that all five implementations return responses accompanied by \texttt{\small RRSIG} records. More importantly, each \texttt{\small NS} referral to the child zone was accompanied by \texttt{\small DS} records. \texttt{\small DNSKEY}s were never included in answers and were, thus, queried separately. We choose BIND9 as an authoritative nameserver to set up our enumeration infrastructure.

\subsection{Recursive Resolvers\label{recursive_resolvers}}

While we can freely choose any suitable authoritative nameserver implementation, we do not have any control over recursive resolvers. We again refer to RFC 4035~\cite{rfc4035} and study the expected behavior of validating recursive DNS resolvers from 5 software vendors, summarized in Table~\ref{validator_properties}. For this experiment, we have deployed the same two-level (\texttt{\small test.com} and \texttt{\small subdomain.test.com}) authoritative DNS zone setup using BIND9 as an authoritative nameserver. We study the three following aspects of the software:

\subsubsection{\texttt{\small DO} bit} A validating recursive resolver must set the \texttt{\small DO} bit on all its requests to signal DNSSEC support, even if the initiating query did not require it. In the tests described below, all the recursive resolver implementations issued DNS requests with the \texttt{\small DO} bit set to 1. Three vendors (BIND9, Unbound, and Knot Resolver) had DNSSEC validation enabled by default, which makes them protected from cache poisoning attacks out of the box.

\begin{table}[t]
    \caption{Test subdomains for active detection of validating open DNS resolvers} 
    \label{domains} 
    \scriptsize
    \centering 
    \setlength{\tabcolsep}{6pt}
    \begin{tabular}{ll}
         \toprule
             \textbf{Subdomain} &
             \textbf{Misconfiguration} \\
         \midrule
             \texttt{\scriptsize valid} & Correctly signed\\
             \texttt{\scriptsize no-ds} & \texttt{\scriptsize DS} record is not published to the parent zone\\
             \texttt{\scriptsize bad-ds} & \texttt{\scriptsize DS} record in the parent zone contains an error\\
             \texttt{\scriptsize no-key} & Public ZSK is not published in the child zone\\
             \texttt{\scriptsize bad-key} & Public ZSK in the child zone contains an error\\
             \texttt{\scriptsize no-rrsig} & \texttt{\scriptsize RRSIG} over \texttt{\scriptsize DNSKEY} RRset is not published in the child zone\\
             \texttt{\scriptsize bad-rrsig} & \texttt{\scriptsize RRSIG} over \texttt{\scriptsize DNSKEY} RRset in the child zone contains an error\\
             \texttt{\scriptsize exp-rrsig} & \texttt{\scriptsize RRSIG} over \texttt{\scriptsize A} RRset in the child zone has expired\\
         \bottomrule
    \end{tabular}
\end{table}

\begin{table*}[t]
    \caption{Organizational distribution of (non)-validating open resolvers}
    \label{asn_open} 
    \scriptsize
    \centering 
    \begin{tabular}{ccllcccccc}
         \toprule
             \multirow{2}{*}{\textbf{Version}} &  \multirow{2}{*}{\textbf{Rank}} &
             \multirow{2}{*}{\textbf{ASN}} & \multirow{2}{*}{\textbf{Organization name}} & 
             \multirow{2}{*}{\textbf{Country code}} & 
             \multirow{2}{*}{\textbf{Total resolvers}} &
             \multicolumn{2}{c}{\textbf{Validators}} &
             \multicolumn{2}{c}{\textbf{Non-validators}} \\
             \cmidrule(lr){7-8}
             \cmidrule(lr){9-10}
             & & & & & & \textbf{Count} & \textbf{Ratio (\%)} & \textbf{Count} & \textbf{Ratio (\%)} \\
         \midrule
             \multirow{5}{*}{IPv4} & 1 & 16276 & OVH & FR & 1,604 & 393 & 24.5 & 1,211 & 75.5 \\
             & 2 & 3462 & Chunghwa Telecom Co., Ltd. & TW & 1,019 & 132 & 13 & 887 & 87.1 \\
             & 3 & 51167 & Contabo GmbH & DE & 618 & 73 & 11.8 & 545 & 88.2 \\
             & 4 & 9318 & SK Broadband Co Ltd & KR & 604 & 187 & 31 & 417 & 69 \\
             & 5 & 24940 & Hetzner Online GmbH & DE & 571 & 109 & 19.1 & 462 & 80.9 \\
             \midrule
             \multirow{5}{*}{IPv6} & 1 & 51167 & Contabo GmbH & DE & 53 & 8 & 15.1 & 45 & 84.9 \\
             & 2 & 16276 & OVH & FR & 49 & 19 & 38.8 & 30 & 61.2 \\
             & 3 & 3209 & Vodafone GmbH & DE & 30 & 0 & 0 & 30 & 100 \\
             & 4 & 20773 & Host Europe GmbH & DE & 28 & 2 & 7.1 & 26 & 92.9 \\
             & 5 & 63949 & Linode, LLC & US & 23 & 20 & 87 & 3 & 13 \\
         \bottomrule
    \end{tabular}
\end{table*}

\subsubsection{Query patterns} When a query with the \texttt{\small DO} bit set arrives at the authoritative nameserver, the response will include additional records as described in Section~\ref{auth_nameserver}. To establish the chain of trust, all five implementations additionally queried for \texttt{\small DNSKEY} records at the parent and child zones. Regular caching rules also apply to DNSSEC-related resource records. However, we have noticed that PowerDNS explicitly queried for the \texttt{\small DS} record at the parent, even though it was provided along with referrals. So, we consider that the presence of \texttt{\small DNSKEY} and \texttt{\small DS} queries at authoritative nameservers indicates the process of DNSSEC validation.

\subsubsection{Misconfigured domains} A correctly operating validator must not only retrieve DNSSEC resource records but also computationally prove that the zone has not been tampered with. To check this, we deployed 7 misconfigured subdomains under \texttt{\small test.com}. Table~\ref{domains} presents all the misconfigurations. The five recursive resolver vendors (BIND9, Unbound, Microsoft DNS, Power DNS, and Knot Resolver) detected all the misconfigured domains.

\section{Step 1: Detection of Validating Open Resolvers \label{active}}

In this section, we present Step~1 of our method in which we actively enumerate IPv4/IPv6 validating open DNS resolvers on the scale of the Internet. We also prepare for Step~2 - we analyze the sequences of \texttt{\small DS} and \texttt{\small DNSKEY} requests on our authoritative nameservers and the responses received on our scanner. This data constitutes our labeled training dataset for a machine learning-based classification approach that we use to identify validating closed resolvers based only on sequences of observed DNSSEC queries at authoritative nameservers in Step~2.

\subsection{Experimental Setup}

\subsubsection{Identifying DNS resolvers} We start with an open resolver scan in which we use our own custom tool developed for sending DNS requests on a large scale. Our targets are the whole routable IPv4 address space~\cite{routeviews} and a list of responsive addresses in the IPv6 space~\cite{hitlist}. The scanner sends an \texttt{\small A} query for the domain name under our control (\texttt{\small test.com}) to each address. To prevent caching, each domain name contains a random string as well as the encoded IP address of the query destination. During the scan, we constantly process requests received on our authoritative nameservers to avoid IP address churn~\cite{wild} later on. We know that we have reached an open resolver if we see an \texttt{\small A} request for our domain. It is possible, however, that the observed source IP address was different from the one of the original query destination. In this case, we refer to the original query destination as a forwarder. However, for further analysis, we keep only non-forwarders. The presence of intermediate recursive resolvers makes the validation process~\cite{rfc4033} more complex - it is impossible to see where exactly DNSSEC validation takes place. 

\begin{table*}[t]
    \caption{Queries to authoritative nameservers from (non)-validating open resolvers} 
    \label{queries_open} 
    \scriptsize
    \centering 
    \setlength{\tabcolsep}{8pt}
    \begin{tabular}{lcccccccc}
         \toprule
             \multirow{3}{*}{\textbf{Queries}} & 
             \multicolumn{4}{c}{\textbf{IPv4}} & \multicolumn{4}{c}{\textbf{IPv6}} \\
             \cmidrule(lr){2-5}
             \cmidrule(lr){6-9}
              & \multicolumn{2}{c}{\textbf{Validators}} & \multicolumn{2}{c}{\textbf{Non-validators}} &  \multicolumn{2}{c}{\textbf{Validators}} & \multicolumn{2}{c}{\textbf{Non-validators}}\\
              \cmidrule(lr){2-3}
              \cmidrule(lr){4-5}
              \cmidrule(lr){6-7}
              \cmidrule(lr){8-9}
              & \textbf{Count} & \textbf{Ratio (\%)} & \textbf{Count} & \textbf{Ratio (\%)} &  \textbf{Count} & \textbf{Ratio (\%)} & \textbf{Count} & \textbf{Ratio (\%)}\\
          \midrule
              \texttt{\footnotesize None} & 19 & 0.3 & 24,253 & 98.1 & - & - & 384 & 96.2 \\
              \texttt{\footnotesize DNSKEY-p}, \texttt{\footnotesize DNSKEY-c}, \texttt{\footnotesize DS-p} & 5,641 & 81.8 & 321 & 1.3 & 164 & 61.7 & 11 & 2.8 \\
              \texttt{\footnotesize DNSKEY-p}, \texttt{\footnotesize DNSKEY-c} & 1,234 & 17.9 & 120 & 0.5 & 100 & 37.6 & 4 & 1 \\
              \texttt{\footnotesize DNSKEY-c}, \texttt{\footnotesize DS-p} & - & - & 13 & 0.05 & - & - & - & - \\
              \texttt{\footnotesize DNSKEY-p} & 1 & 0.01 & 3 & 0.01 & 1 & 0.4 & - & - \\
              \texttt{\footnotesize DNSKEY-c} & 2 & 0.03 & 12 & 0.05 & 1 & 0.4 & - & - \\
          \midrule
                Total & 6,897 & 100 & 24,722 & 100 & 266 & 100 & 399 & 100 \\
         \bottomrule
    \end{tabular}
\end{table*}

\subsubsection{Validation}
Once we obtain an intermediate list of open non-forwarders, we perform a series of queries to determine if they are validators. Each non-forwarder is requested to resolve eight types of subdomains under \texttt{\small dnssec-test.com}, corresponding to each misconfiguration. Each domain name is unique thanks to a random string. To avoid potential packet losses, we repeat each query 10 times. As a result, each resolver is requested to process 80 \texttt{\small A} requests. A correctly configured DNSSEC validating recursive resolver should return to our scanner the \texttt{\small SERVFAIL} error message for misconfigured subdomains, as it cannot complete the validation process. We have verified this correct behavior by querying the Google Public DNS~\cite{google}. Moreover, the two online DNS zone analysis tools (DNSViz~\cite{dnsviz} and DNSSEC Analyzer~\cite{dnssec-analyzer}) interpreted our misconfigurations as intended and marked zones as bogus. If the recursive resolver is DNSSEC-enabled but not validating, it successfully resolves all the tested domain names. 

\subsection{Enumeration Results}

\subsubsection{Open resolver scan}

We scanned 2.8B routable IPv4 and 297M IPv6 addresses from the target list. The set of addresses already excludes the address ranges that we were asked not to scan (see Section~\ref{ethics} for ethical considerations). The scan resulted in 4.6M IPv4 and 5.5K IPv6 open resolvers that contacted our authoritative nameservers to resolve our requests. The majority of them, however, were forwarders and did not perform the recursive resolution themselves. Thus, after the elimination of forwarders, we keep 51.6K IPv4 and 853 IPv6 non-forwarding open resolvers for further analysis.

\subsubsection{DNSSEC-enabled resolvers}

We first examine whether open resolvers declare the DNSSEC support by setting the \texttt{\small DO} bit to 1 in incoming queries for the \texttt{\small valid.dnssec-test.com} subdomain. From the list of non-forwarders derived in the previous step, we further exclude those that did not respond to any of the ten \texttt{\small valid.dnssec-test.com} queries. Neither did we analyze resolvers that this time forwarded their requests. As a result, we obtain 49K IPv4 and 767 IPv6 open non-forwarders. At this step, we analyze at most ten queries per resolver. If a resolver signals its support for DNSSEC in at least one of its requests, we call it DNSSEC-enabled. Such DNSSEC-enabled resolvers are the majority of tested non-forwarders: 38.6K (78.7\%) IPv4 and 710 (92.6\%) IPv6. 

\subsubsection{Validating resolvers}

Validators, apart from being DNSSEC-enabled, are expected to computationally verify the authenticity of received responses. If this check succeeds, they return the response with the \texttt{\small NOERROR} code to the requesting client and resolve the domain to the configured IP address. In case of validation failure, one receives a \texttt{\small SERVFAIL} error. We refer to a DNSSEC-enabled recursive resolver as a \emph{validator} if it never succeeds to resolve any of the misconfigured domain names. Although the DNS return codes are generic and not directly related to DNSSEC, we consider \texttt{\small SERVFAIL} to be the result of the validation failure and \texttt{\small NOERROR} as its success. The \texttt{\small no-ds} subdomain is a special case that we discuss later.

We continue to analyze DNSSEC-enabled open resolvers derived in the previous step. We retrieve the response codes for all the six misconfigured subdomains (excluding \texttt{\small no-ds}) returned to the scanner machine. We further rule out resolvers that did not return any response for one or more subdomains. Neither do we keep those that returned multiple response codes for a single misconfiguration. As a result, 7.3K IPv4 and 273 IPv6 open resolvers successfully detected all our misconfigurations and responded with the \texttt{\small SERVFAIL} return code to all the queries. They represent 18.9\% IPv4 and 38.5\% IPv6 DNSSEC-enabled resolvers.

\subsubsection{Absence of \texttt{\small DS} at the parent zone}

One has to manually add the generated \texttt{\small DS} records to the parent zone, for example through a registrar's control panel. This process is error-prone~\cite{registrars} and the absence of the key hash at the parent zone is one of the most common server-side misconfigurations~\cite{longitudinal}. The DNSSEC standard requires treating the domains without \texttt{\small DS} at the parent zone as if they were unsigned~\cite{rfc4035}, which makes the zone insecure from the DNSSEC point of view but should not result in a validation error. We now check how validating resolvers (that successfully detected all the six misconfigurations) treat our zone without \texttt{\small DS} records. The great majority of them returned \texttt{\small NOERROR} to queries for \texttt{\small no-ds.dnssec-test.com} domains. The remaining 389 IPv4 and 7 IPv6 resolvers failed, so we excluded them from the list of validators. As a result, we identified 6.9K IPv4 (266 IPv6) validators and 24.7K IPv4 (395 IPv6) non-validators.

\subsubsection{Non-validating and partially validating resolvers}

In general, the great majority of DNSSEC-enabled resolvers are consistent - they either validate or invalidate all the misconfigurations. Very few resolvers (43 IPv4 and 4 IPv6) returned different response codes for different subdomains. In the remainder of the paper, we will consider them as \textit{non-validators}. Finally, the majority of DNSSEC-enabled resolvers are non-validators: 24.7K (64\%) IPv4 and 399 (56.2\%) IPv6 open resolvers returned \texttt{\small NOERROR} for all the queries to misconfigured zones. Although these resolvers support  DNSSEC (as they set the \texttt{\small DO} bit in outgoing queries), they do not anyhow benefit from it to validate the received responses.

\subsubsection{Autonomous system and geographical distribution}

The discovered open resolvers are widely spread both geographically and organizationally. They originate from 189 countries and 5.6K autonomous systems. Table~\ref{asn_open} presents the top 5 autonomous systems with the highest total number of discovered validating and non-validating (denoted later on as (non)-validating) IPv4/IPv6 open resolvers. It also provides the ratio of each resolver type per autonomous system. The French and German cloud providers (OVH and Contabo) lead for the number of identified resolvers in both address spaces, however, the majority of them are non-validators. In general, autonomous systems are dominated by non-validators, which is the case for 75.1\% IPv4 and 50.6\% IPv6 ASes. Moreover, autonomous systems tend to be highly consistent: 81\% IPv4 and 87\% IPv6 ASes contain either validators or non-validators, but not both.

\subsubsection{Software versions} We additionally queried all the (non)-validating open resolvers for DNS software versions and obtained results for 9.4K IPv4 and 244 IPv6 resolvers. Whether validators or not, the majority of these resolvers (92\%) are running different versions of BIND. Other (but much less frequently seen) vendors include Unbound and PowerDNS.

\subsection{Machine Learning-Based Classification}

We already identified validating (6.9K IPv4 and 266 IPv6) and non-validating (24.7K IPv4 and 399 IPv4) open resolvers by examining return codes on the scanning machine. To prepare Step 2, in which we target closed resolvers, we need to find a scheme for detecting a validator that does not rely on the response analysis but is solely based on analyzing the queries arriving at our authoritative nameservers. For each (non)-validating resolver, we check whether it sent a \texttt{\small DS} request to the parent zone (denoted as \texttt{\small DS-p} for the \texttt{\small valid.dnssec-test.com} domain) as well as \texttt{\small DNSKEY} requests to both parent and child zones (\texttt{\small DNSKEY-p} for the \texttt{\small dnssec-test.com} domain and \texttt{\small DNSKEY-c} for the \texttt{\small valid.dnssec-test.com} domain). 

Table~\ref{queries_open} presents the number of IPv4/IPv6 resolvers per each combination of queries and the groups they belong to. This data constitutes our labeled training dataset in which the three query codes (\texttt{\small DS-p}, \texttt{\small DNSKEY-p}, and \texttt{\small DNSKEY-c}) are binary input features to our classifier and the desired binary output value is the resolver class: non-validator (0) or validator (1). We use a decision tree machine learning algorithm as it is the most interpretable algorithm for our classification problem. We split the dataset so that 70\% is used for training and 30\% for testing. We perform all model training and analysis using \texttt{scikit-learn} \cite{PedregosaVGMTGBPWDVPCBPD11}.

\begin{figure}[t]
    \centering
    \includegraphics[width=0.5\textwidth]{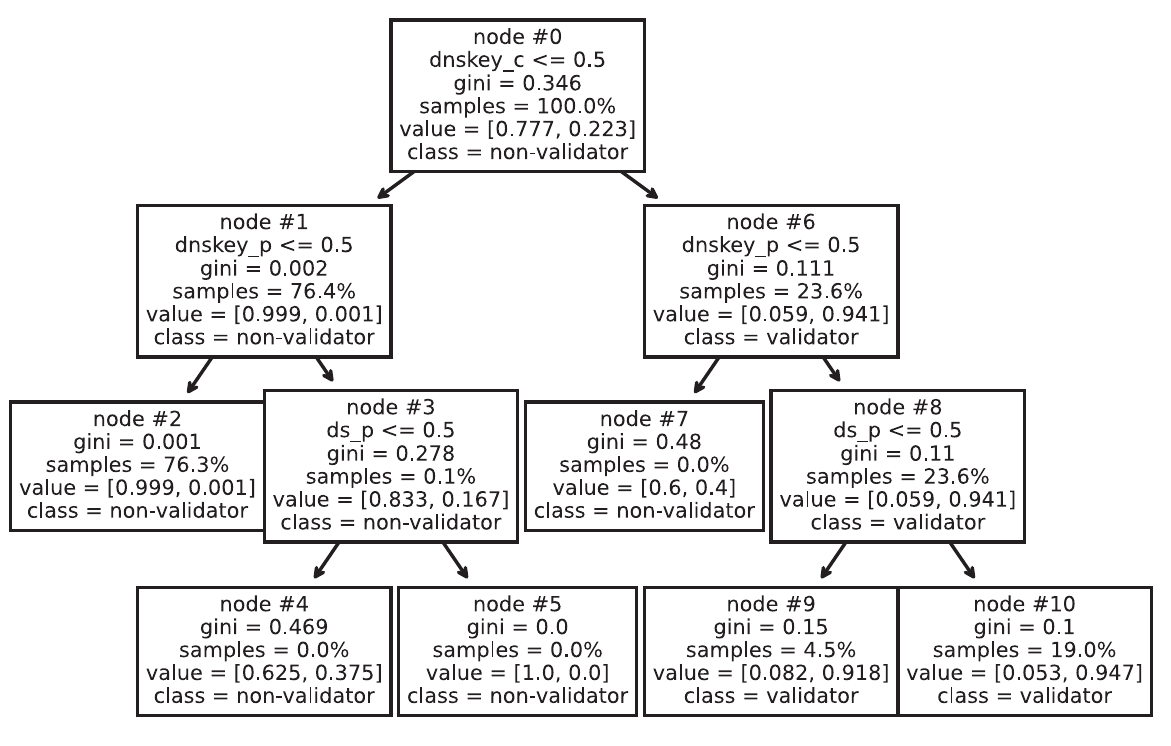}
    \caption{Implementation of the decision tree classifier on a training dataset with three features (\texttt{\small DS-p}, \texttt{\small DNSKEY-p}, and \texttt{\small DNSKEY-c}) and two output classes (validator and non-validator).}
    \label{decision}
\end{figure} 

The proposed method achieves good average weighted results with $Accuracy = 0.984$, $Precision =  0.985 $,  $Recall =  0.984$, $F_{1}-score = 0.984 $, and $MCC = 0.956$. Figure \ref{decision} shows the graphical representation of the decision tree classifier with three input features (query codes \texttt{\small DS-p}, \texttt{\small DNSKEY-p}, and \texttt{\small DNSKEY-c}) and two output classes (validator and non-validator). Each node contains the following information:

\begin{itemize}
    \item Node number -- a unique identifier of the node.
    \item Condition -- applies to non-leaf nodes only. If the condition evaluates to True, the decision process continues to the left child node. If the condition is False, proceed to the right child node. When evaluating the condition, the fest arrow refers to the absence of resource records while the right arrow refers to the presence. 
    \item Gini -- gives the information on whether the items in the dataset belong to different classes. If greater than zero, the sample contains items from different classes.
    \item Samples -- 100\% refers to all the training data. As we go down the tree, the initial training set splits down to samples of smaller sizes.
    \item Value -- a number between 0 and 1. The first item in the list is the proportion of elements at the current node that are non-validators while the second one is the proportion of validators.
    \item Class -- the label assigned to the node. Can be either a validator or a non-validator.
\end{itemize}

We see from Table~\ref{queries_open} that the most observed query pattern was \texttt{\small None}, which means the absence of any queries on our nameservers. In the great majority of cases, it was the characteristic of non-validators (24.3K IPv4 and 384 IPv6). To see how the decision tree labels resolvers that did not issue a single query, we start at the root node. The first condition evaluates to True (no \texttt{\small DNSKEY-c} resource record), so we go to node \#1. The condition at node \#1 also evaluates to True (no \texttt{\small DNSKEY-p} resource record) so we go to the left node \#2. We already finished at the leaf node, although we never checked for the presence of the \texttt{\small DS-p} record. It turns out that the great majority of validators issue at least one of the \texttt{\small DNSKEY} queries, so the classifier labels all resolvers without any \texttt{\small DNSKEY} as non-validators. The data at node \#2 suggests that 99.9\% of resolvers in this group are indeed non-validators while the remaining validators were mistakenly assigned this label.

We have identified 5.9K IPv4 and 175 IPv6 resolvers requesting all the three resource records in zones under our control to establish the chain of trust. As we expect, the great majority of them are validators, an important insight proving that the combination of \texttt{\small DNSKEY-p}, \texttt{\small DNSKEY-c}, and \texttt{\small DS-p} is a strong fingerprint of a validating resolver (see Table~\ref{queries_open} and Figure \ref{decision}, node \#10). Another query pattern largely observed for validators is the transmission of two \texttt{\small DNSKEY} requests (see Figure \ref{decision}, terminal node \#9). As we have shown in Section~\ref{recursive_resolvers}, some validating recursive resolver implementations only request the two keys (\texttt{\small DNSKEY-c} and \texttt{\small DNSKEY-p}) and do not explicitly issue a \texttt{\small DS} query, as they retrieve the resource record directly from the cache. 

Finally, the remaining query patterns were issued by only 31 IPv4 and 2 IPv6 resolvers. Unlike our \texttt{\small A} requests, the \texttt{\small DS} and \texttt{\small DNSKEY} queries do not refer to unique subdomains but rather to zone names. As such, if multiple resolvers share the same cache and one of them already requested the records with valid TTL, other resolvers can retrieve them directly from the cache~\cite{nameserver}, which explains why a few validating resolvers request a subset of necessary records (see Figure \ref{decision}, nodes \#4, \#5, and \#7).

\section{Step 2: Detection of Validating Closed Resolvers\label{passive}}

In this section, we present Step 2 of our method in which we look for closed resolvers by actively probing 3.1B routable IP addresses with packets having forged source addresses. The ultimate goal is to classify them as validators or non-validators using the proposed machine learning-based approach. 

\subsection{Experimental Setup}

We have already examined the behavior of DNSSEC-enabled open resolvers. Now, the goal is to generate queries to our authoritative nameservers on closed resolvers. By definition, they only serve predefined clients and are not reachable from hosts outside. However, a recent study has shown that if a closed resolver receives a request from the same prefix (even when the source IP address of the UDP packet is forged), it is very likely to accept and process the query~\cite{korczyski2020dont}.

For the forged packet to reach the closed resolver, we leverage the absence of Source Address Validation (SAV). The standard requires enabling packet filtering at the network edge and dropping all the traffic with unexpected source IP addresses~\cite{bcp38}. 

\begin{figure}[t]
  \centering
   \includegraphics[width=0.48\textwidth]{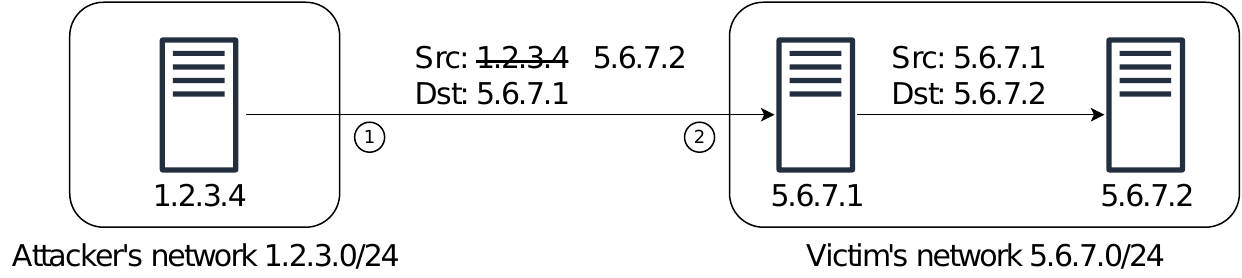}
  \caption{Sending a packet with a forged source IP address.}
  \label{spoofing}
\end{figure}

Figure~\ref{spoofing} shows an example of sending a packet with its source IP forged. SAV can be applied to packets in two directions: outbound or inbound. In the outbound scenario, packets leaving the customer network may have source addresses that do not belong to the prefix of the network. Networks not filtering outgoing traffic can become attack sources. Moreover, even if networks enable ingress filtering~\cite{bcp38}, they will protect the rest of the Internet from their customers but can still be attack targets. As shown in Figure~\ref{spoofing}, the sender transmits a request to host \texttt{\small 5.6.7.1}, but it modifies its IP address to be \texttt{\small 5.6.7.2}. As this address does not belong to the sender prefix (\texttt{\small 1.2.3.0/24}), it should be dropped at the network edge (denoted as \raisebox{.5pt}{\textcircled{\raisebox{-.9pt} {1}}}). In the inbound scenario, packets entering the customer network  have source addresses that belong to the destination network, which reveals to the outsider an otherwise hidden internal part of the network such as closed DNS resolvers. In our example, a packet with the source IP address of \texttt{\small 5.6.7.2} arrives at the \texttt{\small 5.6.7.0/24} network edge (denoted as \raisebox{.5pt}{\textcircled{\raisebox{-.5pt} {2}}}). The packet IP address belongs to the destination network, however, it is not supposed to arrive from the outside and should be dropped.

For our proposed detection of DNSSEC-validating resolvers, we exploit the absence of inbound SAV. We issue \texttt{\small A} requests for the DNSSEC-enabled domain names under our control in a global-scale measurement. We create a two-level DNS zone infrastructure (\texttt{\small test.com} and \texttt{\small subdomain.test.com}) so that we can capture child and parent zone queries. 

The difference with the open resolver scan is that every packet leaving our scanner has a forged source IP address, adjacent to the destination, but always from the same prefix. We rely on the absence of inbound SAV at the destination network edge and in transit networks. If our forged packet freely enters the network and reaches a DNS resolver, we will see the resolution traces on our authoritative nameservers. In parallel to the closed resolver scan, we run an open resolver scan, so that we can later distinguish closed resolvers from the open ones.

\subsection{Enumeration Results}

\subsubsection{Closed and open resolver scans}
Similarly to the open resolver scan described in the previous section, we sent 2.8B IPv4 and 297M IPv6 \texttt{\small A} requests, this time with forged source IP addresses. If a tested resolver only resolves a query seemingly coming from the same prefix, then it is closed. Conversely, a resolver accepting a query from our scanner is open. The measurement campaign took 12 days.

\subsubsection{Identifying DNSSEC-enabled closed resolvers}

Our forged queries reached 5.7M IPv4 and 51.2K IPv6 resolvers in networks not deploying inbound SAV. A majority of those resolvers (92.8\% in IPv4 and 54.3\% in IPv6) are forwarders, so we eliminate them from further analysis. From the remaining non-forwarders, we only keep closed resolvers resulting in 382K IPv4 and 22.9K IPv6 servers. They show a significantly higher level of DNSSEC support compared to open resolvers: 98.8\% IPv4 and 99.2\% IPv6 closed resolvers are DNSSEC-enabled. We keep them for further analysis.

\begin{table}[t]
    \caption{Queries on authoritative nameservers from DNSSEC-enabled closed resolvers} 
    \label{closed_queries} 
    \scriptsize
    \centering 
    \setlength{\tabcolsep}{8pt}
    \begin{tabular}{lcc}
         \toprule
             \textbf{Queries} &  \textbf{IPv4} & \textbf{IPv6} \\
         \midrule
             \texttt{\footnotesize None} & 235,534 & 9,294 \\
             \texttt{\footnotesize DNSKEY-p}, \texttt{\footnotesize DNSKEY-c}, \texttt{\footnotesize DS-p} & 128,374 & 8,422 \\
             \texttt{\footnotesize DNSKEY-p}, \texttt{\footnotesize DNSKEY-c} & 12,897 & 1,335\\
             \texttt{\footnotesize DNSKEY-c}, \texttt{\footnotesize DS-p} & 58 & 2 \\
             \texttt{\footnotesize DNSKEY-p}, \texttt{\footnotesize DS-p} & 79 & 13 \\
             \texttt{\footnotesize DNSKEY-p} & 140 & 5 \\
             \texttt{\footnotesize DNSKEY-c} & 230 & 3,689 \\
             \texttt{\footnotesize DS-p} & 25 & 1 \\
             \midrule
             Total  & 377,337  & 22,761 \\
         \bottomrule
    \end{tabular}
\end{table}

\begin{table*}[t]
    \caption{Organizational distribution of (non)-validating closed resolvers}
    \label{asn_closed} 
    \scriptsize
    \centering 
    \begin{tabular}{cccccccccc}
         \toprule
             \multirow{2}{*}{\textbf{Version}} &  \multirow{2}{*}{\textbf{Rank}} &
             \multirow{2}{*}{\textbf{ASN}} & \multirow{2}{*}{\textbf{Organization name}} & 
             \multirow{2}{*}{\textbf{Country code}} & 
             \multirow{2}{*}{\textbf{Total resolvers}} &
             \multicolumn{2}{c}{\textbf{Validators}} &
             \multicolumn{2}{c}{\textbf{Non-validators}} \\
             \cmidrule(lr){7-8}
             \cmidrule(lr){9-10}
             & & & & & & \textbf{Count} & \textbf{Ratio (\%)} & \textbf{Count} & \textbf{Ratio (\%)} \\
         \midrule
             \multirow{5}{*}{IPv4} & 1 & 46606 & Unified Layer & US & 44,471 & 59 & 0.1 & 44,412 & 99.9 \\
             & 2 & 14061 & DigitalOcean, LLC & US & 29,497 & 15,150 & 51.4 & 14,347 & 48.7 \\
             & 3 & 63949 & Linode, LLC & US & 10,258 & 5,078 & 49.5 & 5,180 & 50.5 \\
             & 4 & 20473 & The Constant Company, LLC & US & 8,981 & 3,359 & 37.4 & 5,622 & 62.6 \\
             & 5 & 34788 & Neue Medien Muennich GmbH & DE & 7,050 & 0 & 0.0 & 7,050 & 100.0  \\
             \midrule
             \multirow{5}{*}{IPv6} & 1 & 63949 & Linode, LLC & US & 3,283 & 2,053 & 62.5 & 1,230 & 37.5 \\
             & 2 & 	197695 & "Domain names registrar REG.RU", Ltd & RU & 2,262 & 841 & 37.2 & 1,421 & 62.8  \\
             & 3 & 20857 & Transip B.V. & NL & 1,664 & 404 & 24.3 & 1,260 & 75.7  \\
             & 4 & 14061 & DigitalOcean, LLC & US & 1,157 & 462 & 39.9 & 695 & 60.1  \\
             & 5 & 197540 & netcup GmbH & DE & 477 & 211 & 44.2 & 266 & 55.8  \\
         \bottomrule
    \end{tabular}
\end{table*}

\subsubsection{Classification}
To identify validators among DNSSEC-enabled closed resolvers, we retrieve the queries on our authoritative nameservers (see the details in Table~\ref{closed_queries}). Similarly to the open resolver scan, the patterns that largely dominate are \texttt{\small DNSKEY-p + DNSKEY-c}, \texttt{\small DNSKEY-p + DNSKEY-c + DS-p}, and the absence of any DNSSEC-related query. In Section~\ref{active}, we have presented a classification model that predicts whether a given resolver is a DNSSEC validator by examining its queries to authoritative nameservers. We run the model on the unseen data and label 37.4\% IPv4 and 42.9\% IPv6 closed resolvers as validators.

\subsubsection{Verification} So far, we found closed validators by inspecting query patterns on our nameservers. To validate the finding, we could have also inspected the DNS response codes but the proposed methodology does not allow us to do so as the responses are sent back to forged source addresses. Instead, we leveraged RIPE Atlas, a global measurement network with probes located in thousands of autonomous systems worldwide~\cite{ripe_atlas}. It lets us both reach closed resolvers (configured as local resolvers of probes) and access DNS response codes. We inspected 287 IPv4 and 71 IPv4 closed resolvers. The ratio of validators was 35.9\% and 42.3\% in IPv4 and IPv6, respectively, consistent with the results returned by the classifier. Moreover, 97 IPv4 and 26 IPv6 resolvers were also tested using our passive methodology, which correctly classified 91\% of those as either validators or non-validators.

\subsubsection{Representativeness} the presented (non)-validators were collected as a result of an active measurement. Previous work showed that responding resolvers can be misconfigured routers or other network equipment rather than operational DNS resolvers~\cite{goingwild}. Our measurements would also trigger otherwise unused systems to respond, which is why we need to evaluate whether the tested resolvers generate any traffic besides our scans.

We have used the Day In The Life of the Internet (DITL) dataset~\cite{ditl} that contains query traces to the DNS root servers gathered during for days. In particular, we have analyzed 301B incoming queries on the DNS root servers from 31M hosts. Importantly, these traces were collected before we started our active measurements, so they do not contain traffic generated by us. Our 432K resolvers (both open and closed) represent 1.39\% of all the resolvers seen at the root servers but accounted for 7.08\% of all the observed traffic. Consequently, we must have measured recursive resolvers that have some real end clients or systems behind them.

\subsubsection{Autonomous System and Geographical Distribution}

The discovered validating and non-validating closed resolvers come from 220 countries and 20K autonomous systems. Table~\ref{asn_closed} presents the organizational distribution and the fraction of (non)-validators. In IPv4, four large American ISPs and telecom companies dominate the rank. Surprisingly, the German hosting provider does not have a single validator but as many as 7K non-validating resolvers. In general, more than half of IPv4 organizations (51.2\%) are predominated by non-validators. This ratio is even larger for IPv6 (61.2\% ASes have more non-validators than validators).

\subsubsection{Software versions} Similarly to open resolvers, closed (non)-validating resolvers also mostly run the BIND9 software, which is the case for 99.9\% of resolvers for which we managed to retrieve the information.

\section{Ethical Considerations\label{ethics}}

Network scanning has become a well-established practice in the research community. We contacted the Institutional Review Board (IRB) of our university and described the research presented in this paper. Our study is out of their scope because it does not involve human subjects. As it is not uncommon that institution IRBs do not evaluate the research involving computer systems only, the measurement community developed its own guidelines for researchers. In particular, Partridge and Allman advocate that every publication should contain ethical analysis so that it can be further discussed in the community~\cite{ethics}. 
The Menlo report, in turn, outlines the recommended research practices in the field of Information and Communications Technologies~\cite{menlo}.

Durumeric et al.~\cite{zmap} provided technical guidelines for performing large-scale active measurements. To minimize the interference caused by our measurements, we have randomized the scanning lists so that we do not consecutively scan all the machines inside the same network at the same time. Moreover, all the scanned domain names point to a website that provides contact details for opting out. We did not scan any hosts that opted out from our previous measurement campaigns and did not receive any new complaints.


Generating traffic with forged IP addresses has been extensively used in research to check for the deployment of ingress filtering~\cite{bb-spoofer-sruti,Beverly:2009:UED:1644893.1644936,spoofer_new,spoofer,casey,korczyski2020dont} and to identify the presence of censorship~\cite{sundararaman2020censoredplanet,Paul2018Towards,Paul2017Augur:,Roya2015Analyzing}, which did not raise 
concerns and the methods used were justified for addressing research questions successfully. We build upon the experience of previous work to conduct our 
measurements.

\section{Conclusions\label{conclusions}}

In this paper, we presented a new active method to enumerate DNSSEC-validating recursive resolvers. It overcomes the limitations of other techniques as it is fast, remote, and free.

In the first phase, we scanned for open non-forwarding DNS resolvers and found that while most of them are DNSSEC-enabled, less than 18\% in IPv4 and 38\% in IPv6 validate received responses. We have acquired the ground truth data by examining the query patterns of open resolvers on our nameservers and the actual validation status determined from DNS query codes. We have trained and tested a decision tree classifier resulting in very high scores. 

In the second phase, we probed for closed resolvers and ran our classifier on the query patterns of closed resolvers observed at our two-level authoritative DNS zone infrastructure. The algorithm has shown that 37.4\% IPv4 and 42.9\% IPv6 closed resolvers are likely to be validators. We cross-checked these findings using probes from the RIPE Atlas network. The ratio of validators among closed local resolvers of probes was consistent with our results and we correctly classified 91\% of resolvers that were common with our passive detection dataset.

Our methodology has a large coverage as we have identified (non)-validators in 224 countries and 21.9K autonomous systems. Moreover, the measured DNS resolvers generated a non-negligible amount of traffic observed on DNS root servers, which suggests that they are in use by real customers or other systems. Consequently, the end users/systems that rely on non-validators for DNS resolution remain vulnerable to cache poisoning attacks. 

\section*{Acknowledgment}

This work has been partially supported by Carnot LSI and Grenoble Alpes Cybersecurity Institute (under the contract ANR-15-IDEX-02), the French Ministry of Research projects PERSYVAL-Lab under contract ANR-11-LABX-0025-01, DiNS under contract ANR-19-CE25-0009-01, and the RIPE NCC Community Projects Fund. We thank DNS-OARC for providing access to the Day in the Life of the Internet (DITL) dataset.

\bibliographystyle{IEEEtran}
\bibliography{references}

\begin{thebibliography}{10}
\providecommand{\url}[1]{#1}
\csname url@samestyle\endcsname
\providecommand{\newblock}{\relax}
\providecommand{\bibinfo}[2]{#2}
\providecommand{\BIBentrySTDinterwordspacing}{\spaceskip=0pt\relax}
\providecommand{\BIBentryALTinterwordstretchfactor}{4}
\providecommand{\BIBentryALTinterwordspacing}{\spaceskip=\fontdimen2\font plus
\BIBentryALTinterwordstretchfactor\fontdimen3\font minus
  \fontdimen4\font\relax}
\providecommand{\BIBforeignlanguage}[2]{{%
\expandafter\ifx\csname l@#1\endcsname\relax
\typeout{** WARNING: IEEEtran.bst: No hyphenation pattern has been}%
\typeout{** loaded for the language `#1'. Using the pattern for}%
\typeout{** the default language instead.}%
\else
\language=\csname l@#1\endcsname
\fi
#2}}
\providecommand{\BIBdecl}{\relax}
\BIBdecl

\bibitem{rfc1035}
P.~Mockapetris, ``{Domain names - implementation and specification},'' RFC
  1035, 1987.

\bibitem{kaminsky}
D.~Kaminsky, ``{It's the End of the Cache as We Know It},'' 2008,
  \url{https://www.slideshare.net/dakami/dmk-bo2-k8}.

\bibitem{cache_injections}
A.~{Klein}, H.~{Shulman}, and M.~{Waidner}, ``{Internet-Wide Study of DNS Cache
  Injections},'' in \emph{IEEE INFOCOM}, 2017.

\bibitem{patched_dns}
A.~Herzberg and H.~Shulman, ``{Security of Patched DNS},'' in \emph{ESORICS},
  2012.

\bibitem{fragmentation_leaking}
H.~Shulman and M.~Waidner, ``{Fragmentation Considered Leaking: Port Inference
  for DNS Poisoning},'' in \emph{ACNS}, 2014.

\bibitem{vulnerable_delegation}
A.~Herzberg and H.~Shulman, ``{Vulnerable Delegation of DNS Resolution},'' in
  \emph{ESORICS}, 2013, pp. 219--236.

\bibitem{socket_overloading}
------, ``{Socket Overloading for Fun and Cache-Poisoning},'' in \emph{ACSAC},
  2013.

\bibitem{fragmentation_poisonous}
{A. Herzberg and H. Shulman}, ``{Fragmentation Considered Poisonous, or:
  One-domain-to-rule-them-all.org},'' in \emph{IEEE CNS}, 2013.

\bibitem{collaborative}
F.~{Alharbi}, J.~{Chang}, Y.~{Zhou}, F.~{Qian}, Z.~{Qian}, and
  N.~{Abu-Ghazaleh}, ``{Collaborative Client-Side DNS Cache Poisoning
  Attack},'' in \emph{IEEE INFOCOM}, 2019.

\bibitem{wrinkle}
H.~{Berger}, A.~{Dvir}, and M.~{Geva}, ``{A Wrinkle in Time: A Case Study in
  DNS Poisoning},'' in \emph{Int. J. Inf. Secur.}, 2019.

\bibitem{corrupted}
D.~Dagon, C.~Lee, W.~Lee, and N.~Provos, ``{Corrupted DNS Resolution Paths: The
  Rise of a Malicious Resolution Authority},'' in \emph{{NDSS}}, 2008.

\bibitem{korczyski2020dont}
M.~Korczyński, Y.~Nosyk, Q.~Lone, M.~Skwarek, B.~Jonglez, and A.~Duda,
  ``{Don't Forget to Lock the Front Door! Inferring the Deployment of Source
  Address Validation of Inbound Traffic},'' in \emph{{PAM}}, 2020.

\bibitem{korczyski2020closed}
Y.~Nosyk, M.~Korczy\'nski, Q.~Lone, M.~Skwarek, B.~Jonglez, and A.~Duda, ``{The
  Closed Resolver Project: Measuring the Deployment of Inbound Source Address
  Validation},'' \emph{IEEE/ACM Transactions on Networking}, 2023.

\bibitem{anrw}
M.~Korczy\'{n}ski, Y.~Nosyk, Q.~Lone, M.~Skwarek, B.~Jonglez, and A.~Duda,
  ``{Inferring the Deployment of Inbound Source Address Validation Using DNS
  Resolvers},'' in \emph{ANRW}, 2020.

\bibitem{manipulation}
P.~Pearce, B.~Jones, F.~Li, R.~Ensafi, N.~Feamster, N.~Weaver, and V.~Paxson,
  ``{Global Measurement of {DNS} Manipulation},'' in \emph{{USENIX Security}},
  2017.

\bibitem{rfc5452}
B.~Hubert and R.~Mook, ``{Measures for Making DNS More Resilient against Forged
  Answers},'' RFC 5452, 2009.

\bibitem{rfc4033}
S.~Rose, M.~Larson, D.~Massey, R.~Austein, and R.~Arends, ``{DNS Security
  Introduction and Requirements},'' RFC 4033, 2005.

\bibitem{rfc4034}
------, ``{Resource Records for the DNS Security Extensions},'' RFC 4034, 2005.

\bibitem{rfc4035}
------, ``{Protocol Modifications for the DNS Security Extensions},'' RFC 4035,
  2005.

\bibitem{longitudinal}
T.~Chung, R.~van Rijswijk-Deij, B.~Chandrasekaran, D.~Choffnes, D.~Levin, B.~M.
  Maggs, A.~Mislove, and C.~Wilson, ``{A Longitudinal, End-to-End View of the
  DNSSEC Ecosystem},'' in \emph{{USENIX Security}}, 2017.

\bibitem{registrars}
T.~Chung, R.~van Rijswijk-Deij, D.~Choffnes, D.~Levin, B.~M. Maggs, A.~Mislove,
  and C.~Wilson, ``{Understanding the Role of Registrars in DNSSEC
  Deployment},'' in \emph{IMC}, 2017.

\bibitem{practical_impact}
W.~Lian, E.~Rescorla, H.~Shacham, and S.~Savage, ``{Measuring the Practical
  Impact of {DNSSEC} Deployment},'' in \emph{USENIX Security}, 2013.

\bibitem{measuring_occurence}
M.~Wander and T.~Weis, ``{Measuring Occurrence of DNSSEC Validation},'' in
  \emph{PAM}, 2013.

\bibitem{check_repeat}
{Yingdi Yu}, D.~{Wessels}, M.~{Larson}, and {Lixia Zhang}, ``{Check-Repeat: A
  New Method of Measuring DNSSEC Validating Resolvers},'' in \emph{IEEE INFOCOM
  Workshops}, 2013.

\bibitem{wild}
{\'O}.~Gu{\dh}mundsson and S.~Crocker, ``{Observing DNSSEC Validation in the
  Wild},'' in \emph{SATIN}, 2011.

\bibitem{counting}
K.~{Fukuda}, S.~{Sato}, and T.~{Mitamura}, ``{A Technique for Counting DNSSEC
  Validators},'' in \emph{IEEE INFOCOM}, 2013.

\bibitem{routeviews}
``{University of Oregon Route Views Project},''
  \url{http://www.routeviews.org/routeviews/}.

\bibitem{hitlist}
O.~Gasser, Q.~Scheitle, P.~Foremski, Q.~Lone, M.~Korczy\'nski, S.~D. Strowes,
  L.~Hendriks, and G.~Carle, ``{Clusters in the Expanse: Understanding and
  Unbiasing IPv6 Hitlists},'' in \emph{IMC}, 2018.

\bibitem{ripe_atlas}
``{RIPE Atlas},'' \url{https://atlas.ripe.net}.

\bibitem{bind}
{Internet Systems Consortium}, ``{BIND9},'' \url{https://www.isc.org/bind/},
  Apr. 2023.

\bibitem{microsoft}
{Microsoft}, ``{Domain Name System (DNS)},''
  \url{https://learn.microsoft.com/en-us/windows-server/networking/dns/dns-top},
  Jul. 2023.

\bibitem{nsd}
{NLnet Labs}, ``{NSD},'' \url{https://www.nlnetlabs.nl/projects/nsd/about/},
  Jul. 2023.

\bibitem{pdns}
{PowerDNS.COM BV}, ``{https://www.powerdns.com},''
  \url{https://www.powerdns.com}, Jul. 2023.

\bibitem{knot-dns}
{cz.nic}, ``{Knot DNS},'' \url{https://www.knot-dns.cz}, Jul. 2023.

\bibitem{knot-resolver}
------, ``{Knot Resolver},'' \url{https://www.knot-resolver.cz}, Jul. 2023.

\bibitem{unbound}
{NLnet Labs}, ``{UNBOUND},''
  \url{https://www.nlnetlabs.nl/projects/unbound/about/}, Jul. 2023.

\bibitem{google}
{Google}, ``{Public DNS},''
  \url{https://developers.google.com/speed/public-dns}, Feb. 2023.

\bibitem{dnsviz}
{DNSViz}, ``{A DNS visualization tool},'' \url{https://dnsviz.net}, Jul. 2023.

\bibitem{dnssec-analyzer}
{VERISIGN Labs}, ``{DNSSEC Analyzer},''
  \url{https://dnssec-analyzer.verisignlabs.com}, Jul. 2023.

\bibitem{PedregosaVGMTGBPWDVPCBPD11}
F.~Pedregosa, G.~Varoquaux, A.~Gramfort, V.~Michel, B.~Thirion, O.~Grisel,
  M.~Blondel, P.~Prettenhofer, R.~Weiss, V.~Dubourg, J.~VanderPlas, A.~Passos,
  D.~Cournapeau, M.~Brucher, M.~Perrot, and E.~Duchesnay, ``{Scikit-Learn:
  Machine Learning in Python},'' \emph{Journal of Machine Learning Research},
  vol.~12, pp. 2825--2830, 2011.

\bibitem{nameserver}
A.~Berger, N.~Weaver, R.~Beverly, and L.~Campbell, ``{Internet Nameserver IPv4
  and IPv6 Address Relationships},'' in \emph{IMC}, 2013.

\bibitem{bcp38}
D.~Senie and P.~Ferguson, ``{Network Ingress Filtering: Defeating Denial of
  Service Attacks which Employ IP Source Address Spoofing},'' RFC 2827, 2000.

\bibitem{goingwild}
M.~K\"{u}hrer, T.~Hupperich, J.~Bushart, C.~Rossow, and T.~Holz, ``{Going Wild:
  Large-Scale Classification of Open DNS Resolvers},'' in \emph{IMC}, 2015.

\bibitem{ditl}
DNS-OARC, ``{DITL},'' \url{https://www.dns-oarc.net/oarc/data/ditl}.

\bibitem{ethics}
C.~Partridge and M.~Allman, ``{Ethical Considerations in Network Measurement
  Papers},'' \emph{Commun. ACM}, vol.~59, no.~10, p. 58–64, Sep. 2016.

\bibitem{menlo}
D.~Dittrich and E.~Kenneally, ``{The Menlo Report: Ethical Principles Guiding
  Information and Communication Technology Research},'' U.S. Department of
  Homeland Security, Tech. Rep., August 2012.

\bibitem{zmap}
Z.~Durumeric, E.~Wustrow, and J.~A. Halderman, ``{ZMap: Fast Internet-wide
  Scanning and Its Security Applications},'' in \emph{USENIX Security}, 2013.

\bibitem{bb-spoofer-sruti}
R.~Beverly and S.~Bauer, ``{The Spoofer Project: Inferring the Extent of Source
  Address Filtering on the Internet},'' in \emph{{USENIX} Steps to Reducing
  Unwanted Traffic on the Internet Workshop}, Jul. 2005.

\bibitem{Beverly:2009:UED:1644893.1644936}
R.~Beverly, A.~Berger, Y.~Hyun, and k.~claffy, ``{Understanding the Efficacy of
  Deployed Internet Source Address Validation Filtering},'' in \emph{IMC},
  2009.

\bibitem{spoofer_new}
M.~Luckie, R.~Beverly, R.~Koga, K.~Keys, J.~Kroll, and k.~claffy, ``{Network
  Hygiene, Incentives, and Regulation: Deployment of Source Address Validation
  in the Internet},'' in \emph{CCS}, 2019.

\bibitem{spoofer}
CAIDA, ``{The Spoofer Project},''
  \url{https://www.caida.org/projects/spoofer/}.

\bibitem{casey}
C.~Deccio, A.~Hilton, M.~Briggs, T.~Avery, and R.~Richardson, ``{Behind Closed
  Doors: A Network Tale of Spoofing, Intrusion, and False DNS Security},'' in
  \emph{IMC}, 2020.

\bibitem{sundararaman2020censoredplanet}
R.~Sundara~Raman, P.~Shenoy, K.~Kohls, and R.~Ensafi, ``{Censored Planet: An
  Internet-Wide, Longitudinal Censorship Observatory},'' in \emph{CCS}, 2020.

\bibitem{Paul2018Towards}
P.~Pearce, R.~Ensafi, F.~Li, N.~Feamster, and V.~Paxson, ``{Towards Continual
  Measurement of Global Network-Level Censorship},'' \emph{IEEE S\&P}, 2018.

\bibitem{Paul2017Augur:}
------, ``{Augur: Internet-Wide Detection of Connectivity Disruptions},'' in
  \emph{IEEE S\&P}, 2017.

\bibitem{Roya2015Analyzing}
R.~Ensafi, P.~Winter, A.~Mueen, and J.~R. Crandall, ``{Analyzing the Great
  Firewall of China Over Space and Time},'' in \emph{PETS}, 2015.

\end{thebibliography}

\end{document}